\documentclass[twocolumn,10pt]{asme2ej}

\usepackage{amssymb}
\usepackage{graphicx}
\usepackage{caption}
\usepackage{subcaption}
\usepackage{amsmath}
\usepackage{xcolor}
\usepackage{algorithm}
\usepackage{algpseudocode}
\usepackage{hyperref}



\usepackage{placeins}
\title{TopoCtrl: Post-Optimization Topology Editing Toward Target Structural Characteristics}
\usepackage{helvet}
\author{Hongrui Chen$^{1}$ \quad Dat Ha $^{2}$  \quad Josephine V. Carstensen$^{2}$\quad Faez Ahmed$^{1}$ \thanks{Address all correspondences to hongrui@mit.edu}
    \affiliation{
        $^1$Department of Mechanical Engineering}
    
    \affiliation{
        $^2$Department of Civil and Environmental Engineering\\
        Massachusetts Institute of Technology\\
        Cambridge, MA, 02139 USA}

}

\begin{document}

\maketitle    

%%%%%%%%%%%%%%%%%%%%%%%%%%%%%%%%%%%%%%%%%%%%%%%%%%%%%%%%%%%%%%%%%%%%%%
\begin{abstract}
{\it Topology optimization can generate high-performance structures, but designers often need to revise the resulting topology in ways that reflect fabrication preferences, structural intuition, or downstream design constraints. In particular, they may wish to explicitly control interpretable structural characteristics such as member thickness, characteristic member length, the number of joints, or the number of members connected to a joint. These quantities are often discrete, non-smooth, or only available through a forward evaluation procedure, making them difficult to impose within conventional optimization pipelines. We present TopoCtrl, a post-optimization control framework that repurposes the latent space of a pre-trained topology foundation model for explicit characteristic-guided editing. Given an optimized topology, TopoCtrl encodes it into the latent space of a latent diffusion model, applies partial noising to preserve instance similarity while creating room for modification, and then performs regression-guided denoising toward a prescribed target characteristic. The core concept is to train a lightweight regression model on latent representations annotated with evaluated structural characteristics, and to use its gradient as a differentiable guidance signal during reverse diffusion. This avoids the need for characteristic-specific reformulations, hand-derived sensitivities, or iterative optimization. Because the method operates through partial noising of an existing topology latent, it preserves overall structural similarity while still enabling characteristic controls. Across representative control tasks involving both continuous and discrete structural characteristics, TopoCtrl produces target-aligned topology modifications while better preserving structural coherence and design intent than indirect parameter tuning or naive geometric post-processing. These results highlight latent diffusion with learned regression guidance as a practical route for explicit and instance-preserving control in topology optimization.
}
\end{abstract}

\section{INTRODUCTION}
\begin{figure*}
\centering

\includegraphics[width=\textwidth]{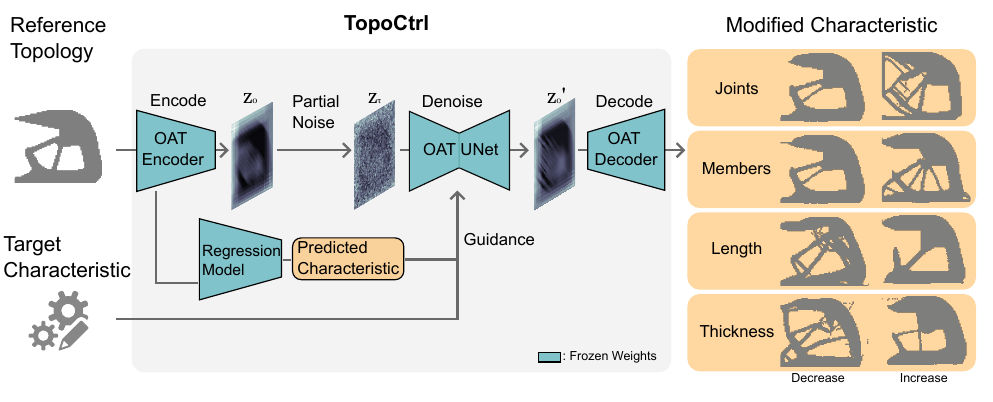}
\caption{TopoCtrl uses a pretrained OAT with frozen weights. Given a reference topology and a user-specified target characteristic (number of joints, maximum members connected to a joint, maximum member length, and member thickness), TopoCtrl encodes the input into OAT’s structured latent, applies partial noising. A small regression model is trained on a small dataset of latent-characteristic pairs to predict the target characteristic. Starting from the noisy latent, we perform diffusion denoising with regression-model guidance toward the target characteristic to recover an edited clean latent that preserves structural similarity. Finally, the edited latent is decoded to yield the edited topology with characteristics matching the target.}

\label{fig:flowchart}
\end{figure*}
Topology optimization has become a powerful paradigm for engineering design, enabling the synthesis of lightweight, high-performance structures through the optimal distribution of material within a prescribed domain. Yet in practical workflows, the optimized result is rarely the end of the design process. Engineers and designers often want to investigate, trust, and revise the solution in ways that reflect downstream constraints, fabrication preferences, and structural intuition. In particular, they may wish to adjust interpretable characteristics of the design, such as member thickness, the number of joints, characteristic member length, or the number of members connected to a joint, while still preserving the overall structural similarity and mechanical performance of the topology. Such requests arise naturally in practice, but they remain difficult to realize within conventional topology optimization pipelines, especially when the desired quantity is discrete, non-smooth, or only available through a forward evaluation procedure. Despite their practical importance, there is still a lack of methods that allow fast, explicit, and instance-preserving control over these challenging structural characteristics.

Existing alternatives partially address this need. One option is to reformulate the optimization problem by introducing additional constraints, penalties, or parameterized controls tailored to the target characteristic \cite{guest2004nodalprojection,guest2009maxlength,lazarov2017maxlengthdensity,fernandez2020minmaxmembersize,zobaer2023maxthickness,allaire2016thicknesscontrol}. While effective for certain quantities, these approaches typically require problem-specific derivations, careful tuning, and repeated optimization solves. The difficulty is even greater for characteristics such as the number of joints or the number of members incident to a joint, which are discrete and not naturally amenable to gradient-based optimization \cite{prager1978finitejoints,kanno2018limitednodes,ohsaki2005nodalstabilityintersection,cai2022trussnodalbuckling,han2021topologicalconstraints2d}. Even varying standard optimization parameters, such as filter size, can introduce some design variability, but this variability is indirect and limited, and it does not amount to explicit control over most of the characteristics that practitioners care about \cite{sigmund2007morphologybw,ha2023humaninformed,schiffer2023hitop2}. Even varying standard optimization parameters, such as filter size, can introduce some design variability, but this variability is indirect and limited, and it does not amount to explicit control over most of the characteristics that practitioners care about. As a result, designers are often left choosing between slow re-optimization, weak indirect control, or ad hoc geometric edits that can easily compromise structural behavior.

Data-driven generative models have recently expanded the design space of topology optimization by learning to predict optimized structures across diverse loading conditions, boundary conditions, and geometric settings \cite{nie2021topologygan,giannone2023diffusing,giannone2023aligning,nobari2025oat}. These methods demonstrate rapid topology generation. However, their primary role is still forward synthesis: given a specification, they generate a candidate topology. They do not directly address the problem of editing an existing optimized design toward a user-specified structural characteristic while preserving its similarity. Furthermore, forward evaluation is relatively easy for many characteristics. Given a topology, one can measure its thickness, count its joints, or extract a characteristic length. The challenge lies in the inverse problem of integrating such evaluations into a controllable design pipeline when only the forward path is readily available. To our knowledge, there is still no general tool that bridges this gap and enables explicit topology control in this setting.

In this work, we address this problem through a regression-guided latent diffusion pipeline, which we call \emph{TopoCtrl}. Figure \ref{fig:flowchart} shows an overview of the proposed method. Our approach builds upon a pre-trained topology foundation model called Optimized Any Topology (OAT) in latent diffusion form and repurposes its structured latent space for controllable post-optimization editing \cite{nobari2025oat,rombach2022latentdiffusion}. Using a relatively small dataset of topologies annotated with evaluated structural characteristics, we train a lightweight regression model that maps the latent representation of the diffusion model to the target characteristic. Once trained, this regressor provides a differentiable guidance signal during denoising, allowing the latent trajectory to be steered toward a desired characteristic value. This approach therefore avoids the need to derive problem-specific sensitivities or reformulate the original topology optimization problem. Instead, we leverage the learned structural knowledge of the diffusion model together with a learned forward predictor of the characteristic to perform explicit control directly in latent space.

Our method relies on partial noising. Rather than sampling a completely new design, we begin from the latent representation of an existing optimized topology and inject only a limited amount of noise before running the reverse diffusion process with regression guidance \cite{song2021ddim,kim2022diffusionclip,kawar2023imagic,mokady2023nulltext,shi2024dragdiffusion,mou2024dragondiffusion}. Partial noising preserves a portion of the original topology information, which helps maintain geometry consistency, while still creating enough freedom for the latent to be manipulated toward the desired target. The subsequent denoising process then allows the foundation model to reconstruct a topology by drawing on the structural knowledge encoded during pre-training \cite{nobari2025oat}. 

The resulting framework is particularly appealing for characteristics that are difficult to differentiate and cumbersome to impose through classical optimization. By decoupling characteristic evaluation from optimization, TopoCtrl opens a practical route for explicit control in settings where only a forward evaluator is available. At the same time, because the method operates on top of a pre-trained foundation model, it inherits broad geometric and physical information from that model and can be applied across structures of varying scales, aspect ratios, and characteristic types.

\noindent Our main contributions are:

\begin{itemize}
\item TopoCtrl: We introduce a characteristic-control pipeline built on a pre-trained latent diffusion topology foundation model, where a lightweight regression model is trained in latent space and used to guide denoising toward prescribed structural characteristics.

\item Explicit control of difficult characteristics: We show that the proposed framework enables explicit control over challenging characteristics, including quantities that are discrete, hard to differentiate, or otherwise inconvenient to integrate into conventional topology optimization formulations.

\item Instance-preserving control through partial noising: We demonstrate that partial noising of an existing topology latent provides an effective mechanism for preserving structural similarity and design intent while still allowing guided modification toward the target characteristic.

\item General applicability across structures and tasks: We show that the latent-guided framework is naturally compatible with structures of different scales and aspect ratios, and can be adapted to different characteristic-control tasks using the same underlying pipeline.
\end{itemize}

The code for this work can be found at: \url{https://github.com/HongRayChen/TopoCtrl}

\section{RELATED WORK}
Our review focuses on data-driven topology optimization, characteristic control in topology optimization, and diffusion-based latent editing and guidance.

\paragraph{Data-driven topology optimization}
Topology optimization distributes material within a design domain to satisfy structural objectives under constraints such as volume fraction. Classical methods include homogenization-based design, shape and topology viewpoints beyond homogenization, material interpolation through Solid Isotropic Material with Penalization (SIMP), and level-set formulations for boundary evolution \cite{bendsoe1988homogenization,rozvany1992generalized,bendsoe1999material,sethian2000structuralboundary,allaire2004sensitivitylevelset}. However, their iterative solve process can be computationally expensive, which motivates learning-based surrogates for rapid generation and exploration.

Data-driven topology optimization addresses this bottleneck by learning a direct mapping from problem specification to a near-optimal structure. Prior work spans conditional GAN and transfer-learning approaches \cite{behzadi2022gantl,sharpe2019cgan}, supervised iteration-free predictors \cite{li2019noniterative,yu2019nearoptimal}, hybrid learning-and-optimization pipelines \cite{zhang2021tonr}, and architectures designed to improve generalization, implicit representation, multiresolution behavior, and three-dimensional generation \cite{wang2022perceptiblecnn,hu2024iftonir,rawat2019conditional,keshavarzzadeh2021ddnsm,zheng2021unet3d}. These methods demonstrate that a dataset of optimized topologies can be learned and used to accelerate topology generation. However, their main role is still the forward synthesis of designs from given boundary conditions and other problem descriptors; they predict an optimized design. They are generally not designed for instance-preserving edits, where the goal is not to generate a new topology from scratch, but to modify one existing design while retaining its structural similarity.

Generative modeling for topology optimization has progressed from GAN-based methods conditioned on physical fields \cite{nie2021topologygan}, to resolution-free implicit generators \cite{nobari2024nito}, and more recently to diffusion-based design generation \cite{maze2023diffusionbeatsgans,giannone2023diffusing,giannone2023aligning}. This progression culminates in OAT, a latent diffusion foundation model trained on a large and diverse corpus of optimized structures across varying boundary conditions, loads, shapes, and resolutions \cite{nobari2025oat}. Compared with earlier data-driven pipelines, such models better capture the broader distribution of optimized designs and provide a structured latent space in which diverse topologies can be represented with a uniform latent tensor. For the present work, this is especially important as explicit characteristic control often pushes a design away from its original optimum, so a strong generative model is needed to reconstruct a topology without significant compromise in compliance. 

\paragraph{Editing and characteristic control in topology optimization}
Although optimized topologies can satisfy performance objectives, they often do not directly satisfy downstream preferences regarding manufacturability, interpretability, or structural style. As a result, a substantial body of work introduces user-facing controls into topology optimization. One class of methods focuses on feature size, thickness, and related geometric descriptors, including interactive feature-size control, graded porosity, and explicit thickness or feature control under density and level-set formulations \cite{ha2023humaninformed,schiffer2023hitop2,schmidt2019gradedporosity,allaire2016thicknesscontrol,guo2014explicitfeature}. Another class introduces appearance-oriented guidance through user-drawn patterns, pattern libraries, patterned shells, multi-pattern constraints, and other designer-in-the-loop mechanisms \cite{schiffer2024interactiveinfill,navez2022patternlibrary,zhu2024holeappearance,meng2023freeformshells,li2024patternshellinfill,zhang2024multipattern,li2023subjective,li2025vrpreference,zhu2025sketchguided,zhang2024sketchaided,mueller2015designerprefs,zhang2026topocnnpattern}. These works show the value of exposing interpretable controls to designers, but they typically do so by reformulating and re-solving the optimization problem itself.

For the characteristics studied in this paper, the literature is more fragmented. Thickness and length scale are among the most extensively studied geometric controls. In density-based and SIMP settings, minimum and maximum member size or length scale are commonly imposed through projection, morphology, filtering, or explicit geometric constraints \cite{poulsen2003minlengthscheme,zhou2001checkerboardmembersize,guest2004nodalprojection,sigmund2007morphologybw,guest2009maxlength,lazarov2017maxlengthdensity,fernandez2020minmaxmembersize,song2024directionalmaxlength,zobaer2023maxthickness}. Related works also pursue direct or explicit thickness and length-scale control, including explicit SIMP-based formulations and discrete-object approaches \cite{zhang2014explicitlengthsimp,zuo2025topologythicknesssimp,carroll2022uniformthicknessdop}. In level-set settings, explicit feature control, skeleton- or boundary-distance-based constraints, uniform thickness formulations, and derivable skeleton representations provide more direct geometric handles \cite{allaire2016thicknesscontrol,guo2014explicitfeature,xia2015boundaryskeleton,wang2016lengthscalelevelset,liu2018uniformthicknesslevelset,tran2022uniformwallthicknessam,huang2024derivableskeletons,wang2025constraintfreevflevelset}. Closely related complexity-control formulations further regulate tunnels, cavities, thinning, and global structural complexity, often with the goal of simplifying or regularizing the topology \cite{han2021topologicalconstraints2d,wang2022topologicalcontrolsimp,zuo2022explicit2dtopcontrol,zuo2023tunnelscavities,liang2022explicitcomplexity2d3d,zhao2020directtopologycontrol,he2022thinningcomplexity,he2023holefillingcomplexity,zhang2017explicitcomplexity,zhang2017mmccomplexity,zhang2016mmcminlength}.

In contrast, controls related to the number of joints or the number of members connected to a joint are much less developed in continuum topology optimization. Related works appear more naturally in truss and layout optimization, where one can directly restrict the number of nodes, simplify connectivity, or impose local nodal stability conditions \cite{prager1978finitejoints,kanno2018limitednodes,fairclough2020simplifiedtrusses,lu2023differentmembers,ohsaki2005nodalstabilityintersection,cai2022trussnodalbuckling}. Some continuum-based topology-control works also indirectly affect structural complexity and connectivity by limiting holes, tunnels, or other topological events \cite{zuo2023tunnelscavities,liang2022explicitcomplexity2d3d,zhao2020directtopologycontrol,he2022thinningcomplexity,he2023holefillingcomplexity}. Nevertheless, explicit post-hoc control over joint count or local joint valence remains difficult. These quantities are often discrete, non-smooth, and only available through a forward evaluation pipeline, making them awkward to embed into conventional gradient-based topology optimization.

These studies highlight both the promise of classical control mechanisms. Many of the differentiable characteristics with explicit controls can already deliver accurate and structurally optimized solutions. If a desired characteristic admits a carefully designed differentiable formulation, one may incorporate it into the optimization problem, but doing so typically requires substantial problem-specific derivation, careful tuning, and repeated iterative solves. Even when the intended edit is local, the optimization update is still global and iterative, and the resulting design may drift from the original instance. While we are not directly competing with existing works on characteristic control, specifically for post-optimization control, we see a limitation: designers may want to preserve the overall similarity of a topology while adjusting one difficult characteristic, rather than solving a new optimization problem from scratch.

\paragraph{Diffusion models, latent editing, and guidance}
Diffusion models learn a data distribution by reversing a gradual noising process \cite{sohldickstein2015nonequilibrium,ho2020ddpm,song2021scorebasedsde}. Practical sampling and control are enabled by accelerated samplers and guidance mechanisms such as Denoising Diffusion Implicit Models (DDIM) and classifier-free guidance \cite{song2021ddim,ho2022classifierfreeguidance}, while strong empirical performance has made diffusion models a dominant generative framework \cite{dhariwal2021diffusionbeatgans}. Latent diffusion improves efficiency by carrying out this process in a compressed latent space rather than directly in pixel space \cite{rombach2022latentdiffusion}, and few-step generation has been further improved through distillation and consistency-based formulations \cite{salimans2022progressivedistillation,song2023consistencymodels,luo2023latentconsistency}.

Diffusion models also demonstrate their success in context aware edits. Prior work demonstrates text- and semantic-guided image editing through attention control, inversion, and instruction-based manipulation \cite{kim2022diffusionclip,hertz2022prompttoprompt,brooks2023instructpix2pix,kawar2023imagic,mokady2023nulltext,cao2023masactrl}. Other works show that sparse interactive constraints, such as point dragging, can be propagated through diffusion features to produce coherent geometric modifications while maintaining instance similarity \cite{zhao2024fastdrag,liu2024dragyournose}. These prior works demonstrated that editing in a learned latent or feature space is often more structurally aware than direct image-space warping, because the model can reconstruct changes using the semantic and geometric information captured during training.

Recent work also shows that diffusion models can be steered by auxiliary predictive models or learned readouts rather than only by text prompts or classifier scores. Plug-and-play guidance, universal guidance, diffusion-feature readouts, and related guidance-by-prediction strategies provide practical mechanisms for controlling the denoising trajectory with external objectives \cite{go2023practicalplugandplay,luo2024readoutguidance,bansal2024universalguidance,kim2022diffusionclip}. This is especially attractive for topology control. In many structural settings, the target characteristic is easy to evaluate in the forward direction once a topology is given, but difficult to differentiate through a conventional optimization pipeline. A learned regression model in latent space offers an alternative that can serve as a lightweight differentiable proxy for the characteristic and provide guidance during denoising without requiring a full reformulation of the original topology optimization problem.

Motivated by these developments, we investigate whether a latent diffusion foundation model for topology optimization can support instance-preserving characteristic control through regression-guided denoising. Compared to direct geometric edits or post-processing, latent-space guidance offers access to the broader structural knowledge learned by the generative model. Relative to reformulating topology optimization with handcrafted characteristic constraints, it avoids characteristic-specific derivations and repeated global solves. This makes it a promising framework for explicit control of challenging structural characteristics, particularly those that are discrete, non-smooth, or otherwise inconvenient to impose in classical optimization.

\section{Proposed Method}
TopoCtrl performs characteristic control by guiding the denoising trajectory of a pre-trained latent diffusion model with a lightweight regression network. Given an optimized topology $T$, we first encode it into the latent space of OAT and apply partial noising so that sampling remains tied to the original structural instance while still permitting controlled variation. We then predict the target structural characteristic from the estimated clean latent and use the resulting gradient to steer reverse diffusion toward a prescribed value. The final latent is decoded back to the topology space by the OAT autoencoder.

\subsection{Diffusion model}
\begin{figure}
\centering
\includegraphics[width=0.45\textwidth]{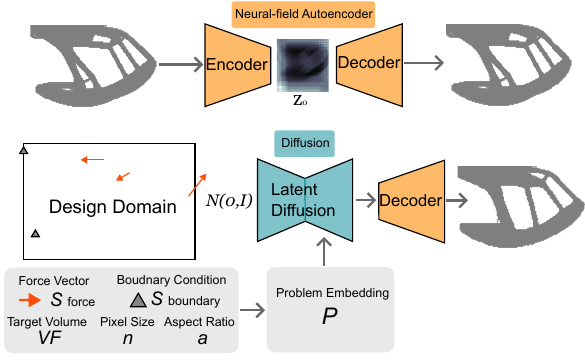}
\caption{An overview of the OAT framework. First, an autoencoder is trained which the encoder encodes problems of arbitrary shapes into a structured fixed resolution latent $z_0$ and the decoder decodes it back to the original topology. Given the problem configuration in the design domain, the problem embedding $P$ is fed into the diffusion model as a conditioning vector. The output from the latent diffusion model is a fixed-resolution latent, which the decoder decodes into a topology.  }
\label{fig:oat_flowchart}
\end{figure}

We build TopoCtrl on top of the pre-trained OAT latent diffusion model~\cite{nobari2025oat}. OAT is well-suited for controllable topology editing because it maps topologies of varying sizes and aspect ratios into a fixed latent tensor, while preserving a neural rendering path that reconstructs the topology at arbitrary resolution. This fixed latent representation makes it convenient to train a single lightweight control model and to perform guidance directly in latent space.

\paragraph{Neural-field autoencoder.}
Let $T \in [0,1]^{H \times W}$ denote an optimized density field defined on a rectangular design domain of height $H$ and width $W$. OAT first encodes $T$ into a latent tensor $z$ through a neural-field autoencoder,
\begin{equation}
z = E(T), \qquad \phi = D(z), \qquad \tilde{T} = R(\phi, c, n),
\end{equation}
where $E(\cdot)$ is the encoder, $D(\cdot)$ is the decoder, and $R(\cdot)$ is a neural renderer. The renderer reconstructs the topology from decoder features $\phi$ using spatial coordinates $c$ and cell sizes $n$, which allows the decoded density field to be queried at arbitrary resolution. The autoencoder is trained with the reconstruction objective
\begin{equation}
\mathcal{L}_{\mathrm{AE}} = \left\| \tilde{T} - T_{\mathrm{GT}} \right\|_1,
\end{equation}
where $T_{\mathrm{GT}}$ denotes the ground-truth topology.

\paragraph{Problem representation.}
Each topology optimization problem is conditioned on the design-domain geometry, boundary conditions, loads, and target volume fraction. Following OAT, we denote the raw problem specification as
\begin{equation}
\hat{P} = \Big(S_{\mathrm{boundary}}, S_{\mathrm{force}}, \mathrm{VF}, n, a \Big),
\end{equation}
where $S_{\mathrm{boundary}}$ is the point set describing support locations and directional constraints, $S_{\mathrm{force}}$ is the point set describing load locations and force vectors, $\mathrm{VF} \in (0,1)$ is the target volume fraction, $n$ encodes the cell size, and $a \in \mathbb{R}^2$ encodes the aspect ratio of the domain. These quantities are embedded into a fixed conditioning vector
\begin{equation}
\begin{aligned}
P = \mathrm{concat}\Big(&\mathrm{BPOM}_{b}(S_{\mathrm{boundary}}), \mathrm{BPOM}_{f}(S_{\mathrm{force}}), \\
&\mathrm{MLP}_{\mathrm{vf}}(\mathrm{VF}),\mathrm{MLP}_{\mathrm{cell}}(n),\mathrm{MLP}_{\mathrm{ratio}}(a)\Big),
\end{aligned}
\end{equation}
which is supplied to the diffusion model during training and inference.

\paragraph{Latent diffusion.}
Let $z_0 = E(T)$ denote the clean latent of the input topology. OAT defines a forward-noising process
\begin{equation}
z_t = \sqrt{\bar{\alpha}_t}\, z_0 + \sqrt{1-\bar{\alpha}_t}\,\epsilon,
\qquad
\epsilon \sim \mathcal{N}(0,I),
\label{eq:topoctrl_forward_noising}
\end{equation}
where $t \in \{0,\dots,T\}$ is the diffusion timestep and $\bar{\alpha}_t$ is the cumulative noise schedule. Rather than predicting $\epsilon$ directly, OAT uses the velocity parameterization
\begin{equation}
v = \sqrt{\bar{\alpha}_t}\,\epsilon - \sqrt{1-\bar{\alpha}_t}\,z_0,
\end{equation}
and trains a conditional UNet $v_\theta$ with the objective
\begin{equation}
\mathcal{L}_{\mathrm{LDM}}
=
\mathbb{E}_{z_0,t,\epsilon}\left[\left\|v - v_\theta(z_t,t \mid \hat{P})\right\|_2^2\right].
\end{equation}

\paragraph{Partial noising and editable sampling.}
TopoCtrl uses partial noising. Instead of starting the reverse process from pure noise, we begin from an intermediate timestep $\tau \ll T$. Given an input topology, we compute its clean latent $z_0$ and form a partially noised latent
\begin{equation}
z_\tau = \sqrt{\bar{\alpha}_\tau}\, z_0 + \sqrt{1-\bar{\alpha}_\tau}\,\epsilon.
\end{equation}
We then run the reverse diffusion process from $\tau$ back to $0$ under regression guidance. Since $\bar{\alpha}_\tau$ remains non-negligible, $z_\tau$ still preserves the dominant semantic content of the original design, including its coarse load path and global layout, while the injected noise creates sufficient flexibility for controlled modification and stochastic sampling of multiple plausible candidates. In contrast, starting from full noise would largely discard instance similarity and turn the process into unconditional regeneration.

\subsection{Regression model}
\paragraph{Medial-axis characteristic evaluation.}
We compute four structural characteristics from each topology using a medial-axis-based post-processing pipeline. Medial-axis skeletonization is a morphological technique used to extract the skeleton of an image. The skeleton consists of all points that are equidistant from the image’s boundary. Starting from a binarized topology, we compute the medial axis and its associated distance field, which together provide a compact representation of the structural centerlines and local thickness. Because raw medial axes are sensitive to small boundary protrusions, we prune short terminal branches with a nominal threshold of 10 pixels. We then identify skeleton junctions and merge nearby thick-region junctions so that visually continuous multi-member joints are counted as a single joint; specifically, candidate joints are merged when their local thickness exceeds 6 pixels and their distance is below 8 pixels. The number of merged joint representatives defines the joint-count characteristic, while the maximum number of incident branches at any merged joint defines the maximum-members-per-joint characteristic. To compute member length, we remove the merged joint regions, split the remaining skeleton into individual members, and take the average length of the two longest members. Because the OAT latent has a fixed spatial size and does not by itself encode the original topology scale, this length is normalized by the longer side of the topology domain. For the thickness characteristic, we use the distance field along the medial axis, convert it to local diameter, and take the 90th percentile value; this quantity is likewise normalized by the longer domain dimension.

\paragraph{Regression model architecture and training.}
Let $y$ denote one of the four target characteristics and let $\tilde{y}$
denote its normalized version. We train four separate regression models, one for each characteristic: number of joints, maximum members going into a joint, average top-two member length, and 90th-percentile thickness. Each regressor takes the latent tensor $z$ as input and outputs a single scalar prediction,
\begin{equation}
\hat{y} = f(z),
\end{equation}
where $f$ is a lightweight convolutional network. In our implementation, $f_\psi$ consists of four convolutional blocks with channel sizes $32$, $64$, $128$, and $256$, each followed by batch normalization and SiLU activation, with spatial downsampling performed by strided convolutions. The resulting feature tensor is globally averaged and passed through a two-layer multilayer perceptron to produce a scalar output. Each regressor is trained independently with mean-squared error,
\begin{equation}
\mathcal{L}_{\mathrm{reg}}=\mathbb{E}_{(z,\tilde{y})}\left[\left\|f(z) - \tilde{y}\right\|_2^2\right].
\label{eq:topoctrl_reg_loss}
\end{equation}

\subsection{Guidance and denoising}
At inference time, the trained regressor is used to guide the denoising toward a prescribed target characteristic. Let $y^\star$ denote the desired characteristic value and let $\tilde{y}^\star$ be its normalized target. Starting from the partially noised latent $z_\tau$, we apply guidance at each reverse step $t=\tau,\tau-1,\dots,1$.

\paragraph{Regression guidance on the estimated clean latent.}
Given the current latent $z_t$, we first query the diffusion model to estimate the corresponding clean latent,
\begin{equation}
\hat{z}_0(z_t,t)=\sqrt{\bar{\alpha}_t}\, z_t-\sqrt{1-\bar{\alpha}_t}\,v_\theta(z_t,t \mid \hat{P}).
\label{eq:topoctrl_x0_pred}
\end{equation}
We then evaluate the regressor on this estimated clean latent and define the control loss
\begin{equation}
\mathcal{L}_{\mathrm{ctrl}}(z_t)=\left\|f_\psi\!\big(\hat{z}_0(z_t,t)\big) - \tilde{y}^\star\right\|_2^2.
\label{eq:topoctrl_ctrl_loss}
\end{equation}
The latent is updated by a gradient step
\begin{equation}
\tilde{z}_t=z_t-\nabla_{z_t}\Big(\lambda_{\mathrm{ctrl}} \, \mathcal{L}_{\mathrm{ctrl}}(z_t)\Big),
\label{eq:topoctrl_guidance_step}
\end{equation}
where $\lambda_{\mathrm{ctrl}}$ controls the strength of characteristic guidance. 

\paragraph{DDIM-style reverse update.}
After the gradient update, we recompute the diffusion prediction on the guided latent $\tilde{z}_t$ and perform a deterministic DDIM reverse step. Specifically, we define
\begin{equation}
\tilde{v}_\theta=v_\theta(\tilde{z}_t,t \mid \hat{P}),
\end{equation}
\begin{equation}
\hat{z}_0(\tilde{z}_t,t)=\sqrt{\bar{\alpha}_t}\,\tilde{z}_t-\sqrt{1-\bar{\alpha}_t}\,\tilde{v}_\theta,
\label{eq:topoctrl_guided_x0}
\end{equation}
and the corresponding noise estimate
\begin{equation}
\hat{\epsilon}(\tilde{z}_t,t)=\sqrt{1-\bar{\alpha}_t}\,\tilde{z}_t+\sqrt{\bar{\alpha}_t}\,\tilde{v}_\theta.
\label{eq:topoctrl_guided_eps}
\end{equation}
The reverse update is then
\begin{equation}
z_{t-1}=\sqrt{\bar{\alpha}_{t-1}}\hat{z}_0(\tilde{z}_t,t)+\sqrt{1-\bar{\alpha}_{t-1}}\hat{\epsilon}(\tilde{z}_t,t).
\label{eq:topoctrl_ddim}
\end{equation}
Repeating this process from $t=\tau$ to $t=0$ yields the final controlled latent, which is decoded by the OAT neural-field autoencoder to produce the edited topology. Because the process begins from a partially noised version of the original instance rather than from pure noise, the resulting design tends to preserve the similarity of the input while moving toward the prescribed characteristic target.

\begin{table}[t]
\footnotesize
\centering
\begin{tabular}{l|c}

Parameter & Value\\
\hline
Regression model training epochs   & 15   \\
Regression training learning rate   & 0.0001   \\
\hline
DDIM Total Steps ($t_{total}$)   & 100   \\
DDIM Partial Steps ($\tau$)  & 30     \\
Guidance strength ($\lambda$)       & 10  \\

\end{tabular}
\caption{Hyperparameter configuration for the topology control tasks. }
\label{tab:hyperparam}
\end{table}

\begin{algorithm}[t]
\caption{Regression-guided latent-based topology control}
\label{alg:topoctrl}
\begin{algorithmic}[1]
\State \textbf{Encode latent:} $z_0 \gets E(T)$
\State \textbf{Add partial noise:} sample $\epsilon \sim \mathcal{N}(0,I)$ and set $z_\tau \gets q_\tau(z_0,\epsilon)$
\State \textbf{Normalize target characteristic:} $\tilde{y}^{\star} \gets \mathrm{Norm}(y^{\star})$
\State Initialize $z_t \gets z_\tau$

\State \textbf{Denoising:} for $t=\tau,\dots,1$ do
\State \hspace{1.4em} Query the diffusion model: $v_\theta(z_t,t \mid \hat{P})$
\State \hspace{1.4em} Estimate the corresponding clean latent:
\State \hspace{2.8em} $\hat{z}_0(z_t,t) \gets \sqrt{\bar{\alpha}_t}\, z_t - \sqrt{1-\bar{\alpha}_t}\, v_\theta(z_t,t \mid \hat{P})$
\State \hspace{1.4em} Evaluate the control loss:
\State \hspace{2.8em} $\mathcal{L}_{\mathrm{ctrl}}(z_t) \gets \left\| f_\psi\!\big(\hat{z}_0(z_t,t)\big) - \tilde{y}^{\star} \right\|_2^2$
\State \hspace{1.4em} Apply regression guidance to the current latent:
\State \hspace{2.8em} $\tilde{z}_t \gets z_t - \nabla_{z_t}\!\left(\lambda_{\mathrm{ctrl}}\, \mathcal{L}_{\mathrm{ctrl}}(z_t)\right)$
\State \hspace{1.4em} Recompute the diffusion prediction on the guided latent:
\State \hspace{2.8em} $\tilde{v}_\theta \gets v_\theta(\tilde{z}_t,t \mid \hat{P})$
\State \hspace{1.4em} Recover the guided clean latent and noise estimate:
\State \hspace{2.8em} $\hat{z}_0(\tilde{z}_t,t) \gets \sqrt{\bar{\alpha}_t}\,\tilde{z}_t - \sqrt{1-\bar{\alpha}_t}\,\tilde{v}_\theta$
\State \hspace{2.8em} $\hat{\epsilon}(\tilde{z}_t,t) \gets \sqrt{1-\bar{\alpha}_t}\,\tilde{z}_t + \sqrt{\bar{\alpha}_t}\,\tilde{v}_\theta$
\State \hspace{1.4em} Take one DDIM reverse step:
\State \hspace{2.8em} $z_{t-1} \gets \sqrt{\bar{\alpha}_{t-1}}\,\hat{z}_0(\tilde{z}_t,t) + \sqrt{1-\bar{\alpha}_{t-1}}\,\hat{\epsilon}(\tilde{z}_t,t)$
\State Denote the final denoised latent by $z_0'$

\State \textbf{Decode structure:} $\phi' \gets D(z_0')$;\quad $T' \gets R(\phi', c, s)$
\end{algorithmic}
\end{algorithm}

We summarize the algorithm in Algorithm \ref{alg:topoctrl}. Table \ref{tab:hyperparam} includes the hyperparameters used for training the regression model and the configurations for partial denoising with characteristics guidance. We use a partial noising to total step ratio similar to existing latent diffusion-based edit and guidance methods \cite{zhao2024fastdrag}.  
\FloatBarrier

\section{RESULTS}
In this section, we set up experiments to demonstrate the characteristic control of the number of joints, the maximum number of members going into a joint, the top two average member lengths, and the 90th percentile thickness. These four characteristics are abbreviated as joints, member, length, and thickness in this section. For the topologies, we select only from the testing dataset of OAT. We conduct the experiment on a PC equipped with an RTX Pro 6000 graphics card and 96GB of VRAM. We generated 64 latent edit results for each of the characteristic control operations for the best-of-N analysis. On average, the inference time on a single topology is around 0.5 seconds per sample.  

For the regression model, we use the pretraining set of the OAT dataset and filter it to a subset of 150K samples for characteristic evaluation and training. This setup is intentionally challenging. The OAT dataset is generated with a fixed filter radius, and for a given boundary condition and volume fraction, the topology optimization solver produces a single deterministic solution. As a result, the training data does not explicitly provide multiple valid solutions with systematically varied characteristics for the same problem instance. Training the regression model for one characteristic takes around 3 minutes. 

For performance evaluation, we use only the test set of OAT. This means that neither OAT nor the regression model saw the test set during training. We select 100 topologies and run characteristic control on each of them. Across the four control tasks, we evaluate 16 target settings in total. For every topology-target pair, we generate 64 stochastic samples from the diffusion model and report both qualitative examples and best-of-64 quantitative results. 

\subsection{Regression model training}
\begin{figure}
\centering
\includegraphics[width=0.4\textwidth]{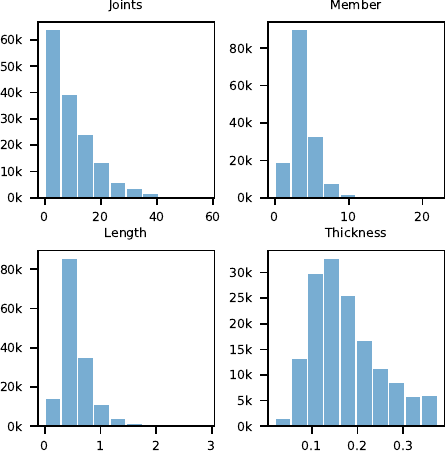}
\caption{The characteristic distributions of the filtered 150K-sample training dataset used for regression model training. We show the four quantities used in TopoCtrl: the number of joints, the maximum number of members incident to a joint, the normalized average of the top two longest members, and the normalized 90th-percentile thickness. The dataset is moderately imbalanced, with joint counts concentrated toward the lower end, most joint complexities lying between 3 and 5, characteristic length concentrated around intermediate values, and thickness spread more broadly across the range. }
\label{fig:ds_char_dist}
\end{figure}

\begin{figure}
\centering
\includegraphics[width=0.4\textwidth]{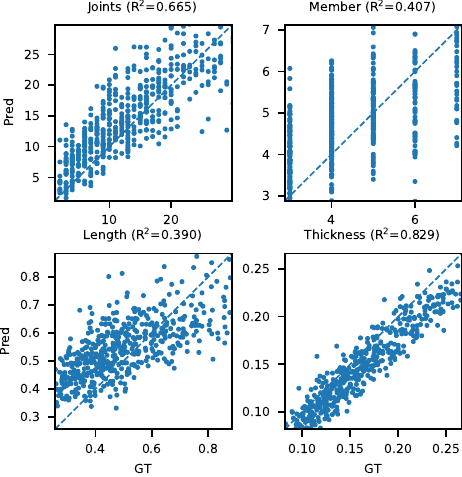}
\caption{Prediction performance of the latent-space regression models on the testing set. Each plot compares the ground-truth characteristic computed from the topology against the value predicted from the latent representation. The regressors recover the main trend of each characteristic. The continuous quantities, namely characteristic length and thickness, exhibit smoother prediction behavior, while the discrete quantities, such as joint count and maximum joint complexity, are more challenging near the tails of the distribution. The length prediction showed slightly higher error around the lower and higher ends of the characteristic values. This also caused the subsequent slight increase in error with the length control evaluations. }
\label{fig:reg_r2}
\end{figure}

We begin by examining the characteristic distribution of the filtered 150K-sample training set. Figure \ref{fig:ds_char_dist} shows the dataset statistics for the four controlled quantities. We observe that the number of joints is concentrated toward the lower end of the range, indicating that highly articulated structures are comparatively rare in the data. The maximum number of members incident to a joint is most often between 3 and 5, which is consistent with the dominant junction patterns observed in the OAT topologies. The normalized average of the two longest members is concentrated around 0.5 to 0.75, while the normalized 90th-percentile member thickness is more broadly distributed. These distributions confirm that the regression task is moderately imbalanced, especially for the more extreme structural cases.

Figure \ref{fig:reg_r2} shows the prediction performance of the latent-space regression models on the held-out test examples. The regressors capture the characteristic trends well enough to serve as useful guidance signals during denoising. The continuous quantities, namely characteristic length and thickness, exhibit smoother prediction behavior, while the discrete quantities, such as joint count and maximum joint complexity, are more challenging near the tails of the distribution. Nevertheless, the results indicate that a relatively small regression model can extract characteristic information from the latent representation of the pre-trained diffusion model. 

\subsection{Characteristic control}
\begin{figure*}
\centering
\includegraphics[width=0.9\textwidth]{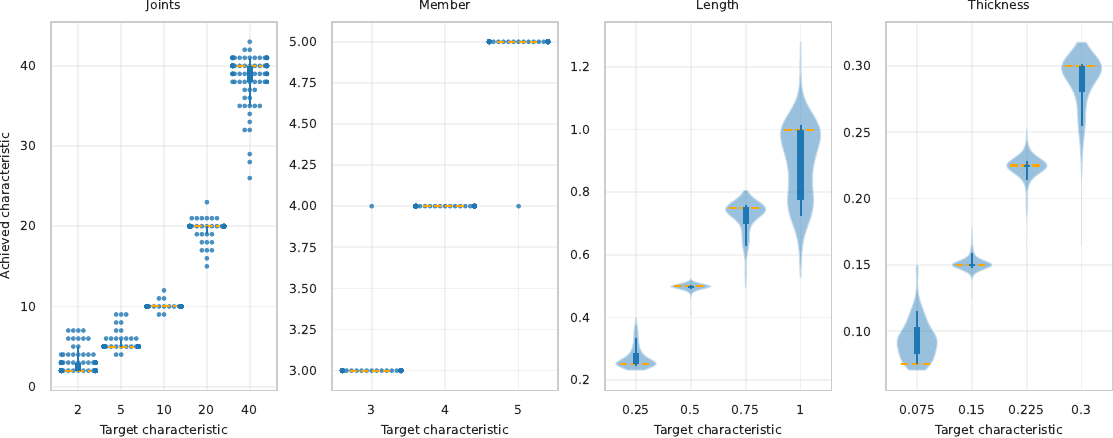}
\caption{Characteristic control accuracy of TopoCtrl over the 100 test topologies using best-of-64 sampling. For the discrete-valued characteristics: the number of joints and the maximum number of members incident to a joint, we use swarm plots. For the continuous-valued characteristics: the normalized characteristic length and normalized 90th-percentile thickness, we use violin plots. In each case, the selected result is the sample whose evaluated characteristic is closest to the requested target under the same medial-axis-based analysis used for dataset annotation. TopoCtrl achieves good target accuracy, with somewhat reduced precision at more extreme target values.}
\label{fig:control_accu}
\end{figure*}

\begin{table*}[t]
\footnotesize
\centering
\setlength{\tabcolsep}{3.5pt}
\renewcommand{\arraystretch}{1.05}
\resizebox{\textwidth}{!}{%
\begin{tabular}{l|cccc|cccc|cccc|cccc}
%\hline
      & \multicolumn{4}{c|}{Joints} & \multicolumn{4}{c|}{Member} & \multicolumn{4}{c|}{Length} & \multicolumn{4}{c}{Thickness} \\
Best of
      & Coverage\% & VFE\% & IoU\% & Failure Rate\%
      & Coverage\% & VFE\% & IoU\% & Failure Rate\%
      & Coverage\% & VFE\% & IoU\% & Failure Rate\%
      & Coverage\% & VFE\% & IoU\% & Failure Rate\% \\
\hline
2   & 66.63 & 0.75 & 76.95 & 45.40 & 64.00 & 0.68 & 80.09 & 38.67 & 48.69 & 0.74 & 77.83 & 43.50 & 68.81 & 1.27 & 74.25 & 49.00 \\
4   & 76.21 & 0.44 & 78.82 & 34.20 & 82.50 & 0.40 & 81.68 & 27.67 & 58.62 & 0.42 & 79.42 & 32.00 & 74.05 & 0.78 & 75.74 & 38.00 \\
8   & 83.50 & 0.24 & 80.18 & 21.60 & 91.00 & 0.23 & 82.85 & 17.67 & 66.31 & 0.23 & 80.41 & 21.50 & 77.42 & 0.51 & 76.89 & 29.00 \\
16  & 87.89 & 0.12 & 81.07 & 15.80 & 98.00 & 0.11 & 83.66 & 10.33 & 72.60 & 0.12 & 81.15 & 15.00 & 80.13 & 0.34 & 77.71 & 22.00 \\
32  & 91.76 & 0.06 & 81.81 & 11.20 & 99.00 & 0.06 & 84.25 &  6.00 & 78.05 & 0.06 & 81.92 &  9.75 & 83.05 & 0.18 & 78.56 & 19.00 \\
64  & 94.82 & 0.03 & 82.51 &  7.40 & 99.50 & 0.03 & 84.93 &  4.67 & 83.84 & 0.03 & 82.55 &  7.00 & 85.02 & 0.10 & 79.21 & 16.00 \\
BM  & 23.89 & 0.21 & 76.45 & N/A   & 56.00 & 0.21 & 76.45 & N/A   & 43.80 & 0.21 & 76.45 & N/A   & 16.94 & 0.21 & 76.45 & N/A   \\
%\hline
\end{tabular}%
}
\caption{Best-of-N quantitative results for TopoCtrl across the four characteristic-control tasks on the testing set. We report coverage, volume-fraction error (VFE), IoU, and failure rate for N=2,4,8,16,32,64, together with the benchmark method obtained by rerunning topology optimization with varying filter radius. Coverage measures how much of the prescribed target interval is actually reached by the method. The failure rate is defined as the percentage of edited topologies whose compliance exceeds two times that of the corresponding ground-truth topology, following the evaluation criterion used in OAT. TopoCtrl achieves higher target coverage than the benchmark while maintaining good IoU, indicating that the generated results remain close to the original instance and function effectively as characteristic-preserving edits.}
\label{tab:bon_all}
\end{table*}

We next evaluate whether the trained regressors can guide the reverse diffusion process toward prescribed target characteristics. For each of the 100 test topologies, we generate 64 samples per target setting. The final characteristic is evaluated using the same geometric analysis used for dataset annotation. In particular, the structural quantities are recomputed from the decoded topology using medial-axis skeletonization and the associated post-processing pipeline, ensuring that training and evaluation are consistent.

Figure \ref{fig:control_accu} summarizes the best-of-64 characteristic accuracy across the four tasks. For the discrete-valued controls, namely the number of joints and the maximum number of members incident to a joint, we use swarm plots. For the continuous controls, including normalized characteristic length and normalized 90th-percentile thickness, we use violin plots. In all cases, the best sample is selected from the 64 stochastic generations according to the smallest absolute deviation from the target value under the same medial-axis skeletonization with identical post-processing settings. TopoCtrl achieves good target accuracy, although performance degrades somewhat at the most extreme target values, such as very high joint counts, the largest target length, or highly extreme thickness values. This behavior is expected, since those cases are both rarer in the data and farther from the nominal optimum represented by the original topology.

\begin{figure}
\centering
\includegraphics[width=0.45\textwidth]{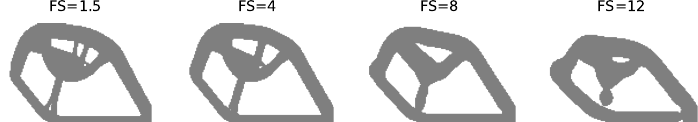}
\caption{Sample benchmark result from the testing dataset rerun with varying filter size (FS). With an increase in filter size, the members become generally thicker, demonstrating some effect in characteristic control. }
\label{fig:fsbm}
\end{figure}

As a benchmark, we also generate comparison topologies by rerunning the testing set of OAT with varying filter radii. For each topology in the test set, we varied the filter radius from 1.5 to 12 with 7 reruns to obtain a variety of solutions (Figre \ref{fig:fsbm}), similar to previous papers in interactive edits \cite{schiffer2023hitop2}. This provides a useful baseline for the amount of characteristic variation that can be obtained through a simple modification of the conventional topology optimization pipeline. Varying the filter radius directly changes feature scale and can indirectly influence topology complexity. However, it remains an indirect mechanism and does not provide explicit instance-level control over most of the evaluated characteristics.

To quantify the reachable target span, we report the coverage metric in Table \ref{tab:bon_all}. Let the prescribed target interval for a given characteristic be $[c_{\min}, c_{\max}]$, and let the interval actually reached by a method over the evaluated results be $[\hat{c}_{\min}, \hat{c}_{\max}]$. We define coverage as
\begin{equation}
\mathrm{Coverage} =\frac{\max\!\left(0,\min(\hat{c}_{\max},c_{\max})-max(\hat{c}_{\min},c_{\min})\right)}{c_{\max}-c_{\min}}.
\label{eq:coverage}
\end{equation}
For example, if the target number of joints spans $[2,40]$ and a method only reaches $[10,20]$, the resulting coverage is $10/38$. Using this metric, we observe that TopoCtrl achieves substantially higher coverage than the baseline obtained by changing the filter radius alone. This is particularly important for the discrete controls, where indirect solver modifications provide only limited variation.

In addition to characteristic accuracy, Table \ref{tab:bon_all} reports compliance, Intersection over Union (IoU), volume-fraction error, and failure rate. Following the original OAT evaluation protocol, we count a result as a failure if its compliance exceeds two times that of the ground-truth topology under the same loading and boundary conditions. TopoCtrl exhibits a somewhat higher failure rate than the original OAT, which is expected because characteristic-controlled topologies are intentionally pushed away from the optimal solution for stiffness. Even so, the IoU remains high across tasks, indicating that the generated outputs remain strongly tied to the original topology and function effectively as edits.

\subsubsection{Joint control}
\begin{figure}
\centering
\includegraphics[width=0.45\textwidth]{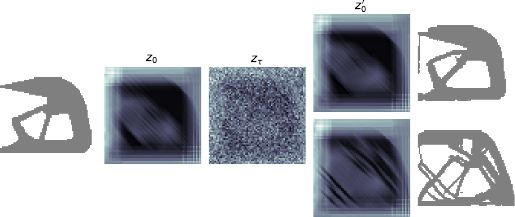}
\caption{TopoCtrl latents for controlling the number of joints. The topology is first converted into latent $z_0$, then partial noise is added to obtain 
$z_\tau$, after which regression-guided denoising is applied to steer the latent toward a lower or higher target joint count, producing the edited latent $z_0'$
. Decoding $z_0'$ gives the controlled topology. The visualization shows how the latent trajectory preserves the overall structural similarity while still allowing the diffusion model to reorganize the connectivity toward the requested level of joint complexity.}
\label{fig:joints_latent}
\end{figure}

\begin{figure*}
\centering
\includegraphics[width=0.7\textwidth]{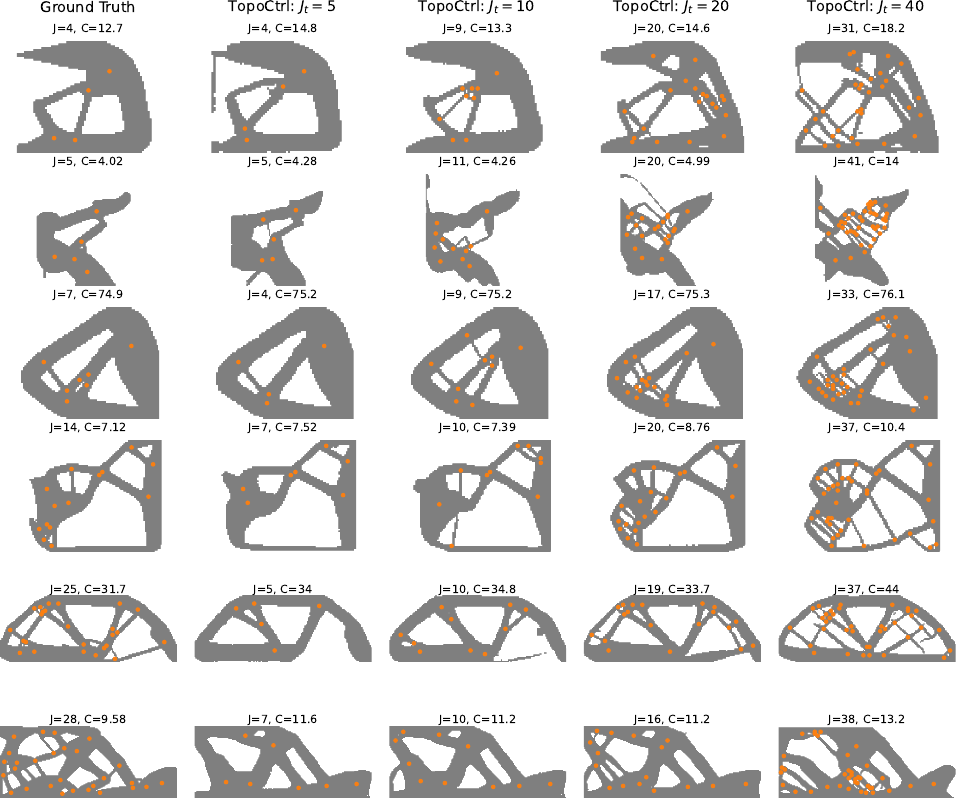}
\caption{Selected examples of joint-count control using TopoCtrl. We define a joint where the medial axis skeletonization branches out. We show six reference topologies together with edited results for target joint counts of 5, 10, 20, and 40. The method successfully steers the topology toward simpler or more articulated structures depending on the target. With lower target values, the generated topologies generally contain fewer branches and connection points. With higher target values, the structures become more complex and develop richer internal connectivity, although reaching very high joint counts can be challenging for initially simple designs. The actual structure generated from TopoCtrl may contain more or fewer joints than the desired target, but in general, they center around the target joints.}
\label{fig:joints_control}
\end{figure*}

We first consider control over the number of joints. Figure \ref{fig:joints_latent} visualizes the latent trajectory for representative examples: the ground-truth topology, the encoded latent $z_0$, the partially noised latent $z_\tau$, the denoised guided latent $z_0'$, and the decoded controlled topologies corresponding to lower and higher target values. Figure \ref{fig:joints_control} then shows six qualitative examples selected from the result set for target joint counts of 5, 10, 20, and 40.

The number of joints can be effectively steered by the regression-guided denoising process. This effect is visible across topologies of varying initial complexity. When fewer joints are requested, the generated topologies generally become simpler, with reduced branching. When more joints are requested, the structures typically become more articulated and develop more complex internal connectivity. Reaching very high targets, such as 40 joints, is more challenging for initially simple structures, but the trend remains clear. 

\subsubsection{Member joint complexity control}
\begin{figure}
\centering
\includegraphics[width=0.45\textwidth]{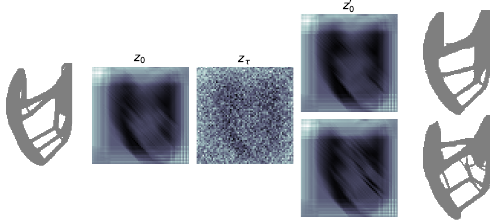}
\caption{TopoCtrl latents for controlling the maximum number of members going into a joint.}
\label{fig:member_latent}
\end{figure}

\begin{figure*}
\centering
\includegraphics[width=0.85\textwidth]{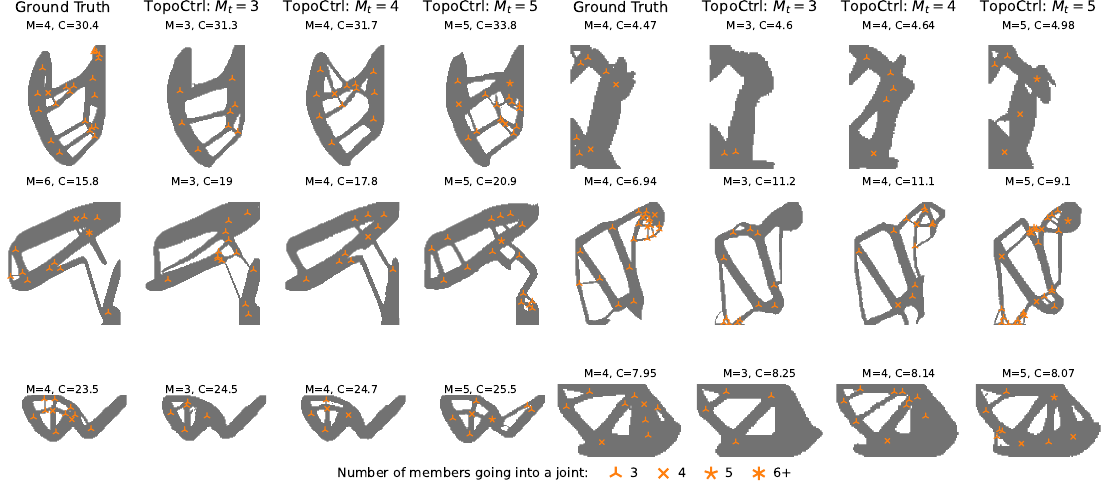}
\caption{Selected examples of the maximum number of members going into a joint control. We show six topologies edited toward target values of 3, 4, and 5 for the maximum number of members connected to a joint. Similar to joint-count control, lower target values encourage simpler connectivity, while higher target values lead to thicker and more complex junction with more members meeting at the same node. }
\label{fig:member_control}
\end{figure*}

We next control the maximum number of members going into a joint. Figure \ref{fig:member_latent} shows representative latent trajectories, and Figure \ref{fig:member_control} visualizes six examples with target values of 3, 4, and 5.

The behavior is qualitatively similar to joint-count control, but now the modification is more localized to the structure of the junctions. When the target is 3, the resulting topologies tend to favor simpler Y-like or chain-like connectivity with fewer highly connected nodes. When the target is increased to 5, the model produces thicker and more complex junction regions with more members meeting at the same node. This control is noteworthy because it targets a discrete and difficult-to-differentiate structural quantity that is cumbersome to impose directly in conventional topology optimization.

\subsubsection{Length control}
\begin{figure}
\centering
\includegraphics[width=0.45\textwidth]{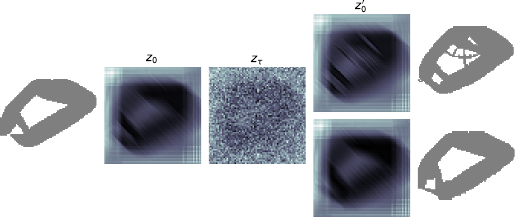}
\caption{TopoCtrl latents for controlling normalized characteristic length, defined as the normalized average of the two longest members extracted from the topology skeleton.}
\label{fig:length_latent}
\end{figure}

\begin{figure*}
\centering
\includegraphics[width=0.7\textwidth]{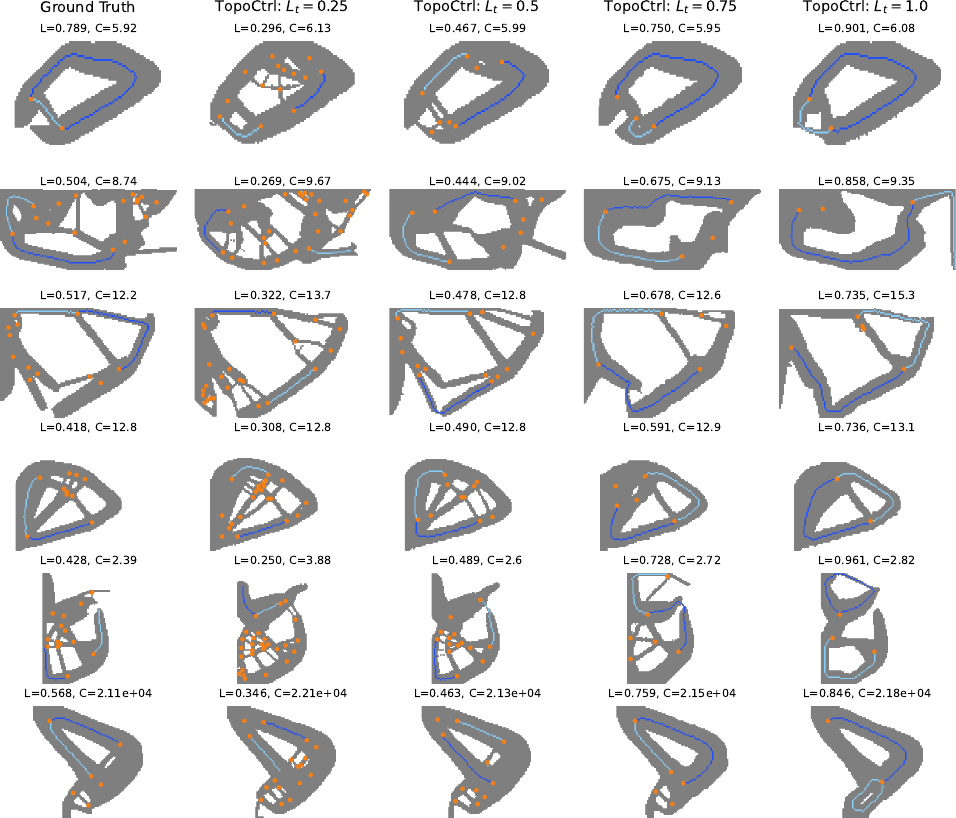}
\caption{Selected examples of characteristic-length control using TopoCtrl. We show six topologies edited toward target values of 0.25, 0.5, 0.75, and 1.0 for the normalized average length of the two longest members. We highlight the longest member in dark blue and the second longest member in light blue. Lower target values generally produce structures with shorter members, while higher target values encourage the formation of longer continuous loops or members.}
\label{fig:length_control}
\end{figure*}

We next study control of normalized characteristic length, defined here as the normalized average of the two longest members extracted from the skeletonized topology. Figure \ref{fig:length_latent} visualizes the corresponding latent trajectories, and Figure \ref{fig:length_control} shows six examples with target values of 0.25, 0.5, 0.74, and 1.0.

Despite the fact that explicit feature-length variation is not part of the original OAT dataset, the regression-guided denoising process still succeeds in steering the topology toward shorter or longer dominant members. Lower target values lead to structures with more fragmented or interrupted long paths, while higher target values encourage the formation of longer continuous loops or members. In particular, for the largest target value of 1.0, we observe generated topologies with visibly elongated structural loops that increase the characteristic member length. This behavior is especially interesting because classical feature-scale control in topology optimization is typically framed as limiting the largest admissible member size, whereas TopoCtrl can also move the design in the opposite direction and deliberately increase characteristic member length.

\subsubsection{Thickness control}
\begin{figure}
\centering
\includegraphics[width=0.45\textwidth]{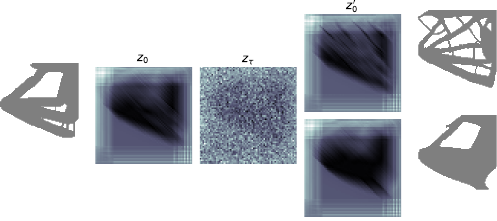}
\caption{TopoCtrl latents for controlling normalized member thickness, measured here as the normalized 90th-percentile thickness of the topology.}
\label{fig:thickness_latent}
\end{figure}

\begin{figure*}
\centering
\includegraphics[width=0.7\textwidth]{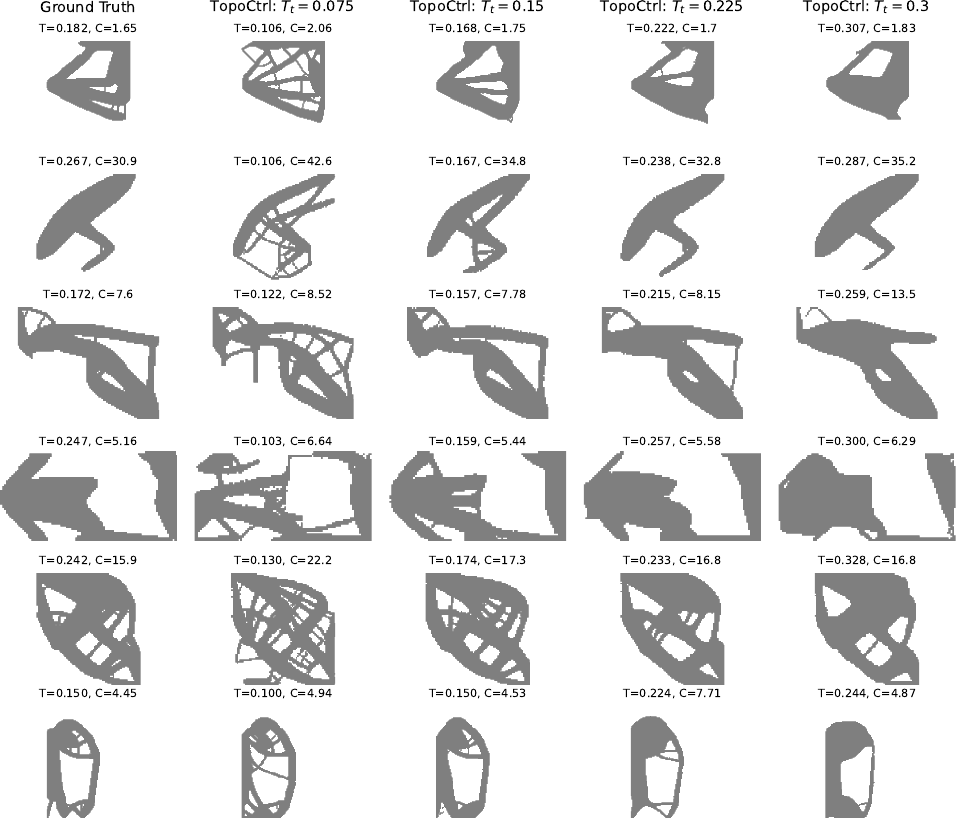}
\caption{Selected examples of thickness control using TopoCtrl. We show six topologies edited toward normalized thickness values of 0.075, 0.15, 0.225, and 0.3. When the target thickness is reduced, the resulting topologies tend to contain thinner members and more branching. When the target thickness is increased, the dominant load-bearing members become thicker.}
\label{fig:thickness_control}
\end{figure*}

Finally, we consider control of normalized member thickness, measured as the normalized 90th-percentile thickness of the topology. Since the characteristic is evaluated at the 90th-percentile, we can still get some members with thicker and thinner than the threshold. Figure \ref{fig:thickness_latent} shows representative latent trajectories, and Figure \ref{fig:thickness_control} presents six examples with target values of 0.075, 0.15, 0.225, and 0.3.

Although the OAT training data is produced with a fixed filter size, TopoCtrl is still able to vary the thickness in both directions. When the target thickness is reduced, the generated topologies tend to contain thinner members and more branching. When the target thickness is increased, the structures become visibly thicker. This result shows that the proposed guidance mechanism can extrapolate beyond the nominal feature scale implied by the original dataset generation process.

\subsection{Cantilever beam }
\begin{figure*}
\centering
\includegraphics[width=0.7\textwidth]{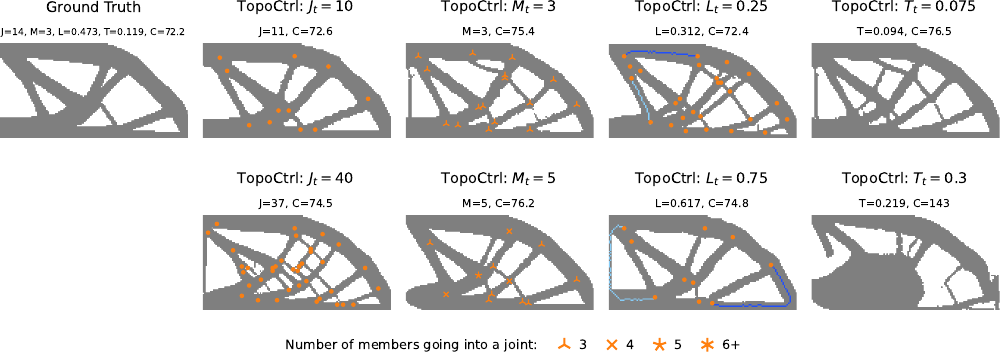}
\caption{Characteristic control of a cantilever beam. The characteristics and compliance of the ground truth topology are evaluated. Lower and higher bound control is then established for the four characteristics. The lower range characteristics are shown in the top row, and the higher range characteristics in the bottom row. Successful control of all four characteristics is observed in the cantilever beam example.  }
\label{fig:cbeam_control}
\end{figure*}
Despite all the previous topologies experimented with being from the testing set, which neither OAT nor the regression model saw during training, generally, the topologies are still within the overall randomized boundary condition style. To verify TopoCtrl’s performance on a common, out-of-distribution topology-optimization test case, we use a cantilever-beam example and control its characteristics (Figure \ref{fig:cbeam_control}). We observe successful characteristic control across all four characteristics.  

These experiments show that TopoCtrl can transform a pre-trained topology foundation model from a forward generator into a control tool. By combining partial noising with regression-guided denoising, the method enables explicit manipulation of structural characteristics that are discrete, non-smooth, or otherwise difficult to integrate into conventional optimization formulations. 

\section{LIMITATIONS}

This work has several limitations that warrant further investigation. First, we have not exhaustively explored all possible structural characteristics, and the current characteristics rely on manually chosen post-processing values. In addition, we did not obtain manual labeling for these characteristics, which may further provide more intuitive labeling. For some characteristics, the regression models also show room for improvement in predictive accuracy. Furthermore, due to the stochastic nature of diffusion models, we relied on up to 64 samples for a single control instance to obtain reliable estimates; with fewer samples, the accuracy remains insufficient, limiting the method’s usefulness as a highly responsive interactive tool for practitioners. Finally, although we demonstrate the approach on both the testing set and a cantilever beam example, it has not yet been extensively evaluated across the full range of common topology optimization problems. As a result, the generalizability of the proposed approach remains to be thoroughly explored.

\section{CONCLUSION AND FUTURE WORK}

In this work, we introduced TopoCtrl, a regression-guided latent diffusion framework for post-optimization control of structural characteristics in topology optimization. Across the experiments, we demonstrated control over four representative characteristics. These results show that TopoCtrl can operate not only on continuous geometric characteristics but also on discrete and difficult-to-differentiate characteristics that are cumbersome to impose through conventional topology optimization pipelines. The method achieved good target-reaching behavior, with broader reachable characteristic coverage than the benchmark based on varying the filter radius, while still maintaining good similarity to the original topology. 
The results highlight the main advantage of TopoCtrl as it provides a practical route to explicit characteristic control in settings where only a forward evaluation procedure is readily available. 

Several directions can extend the present work. First, the four characteristics studied here represent only a small initial set of continuous and discrete quantities chosen to demonstrate the effectiveness of the framework. The same underlying pipeline can be applied to many other structural characteristics, with the practical limitation being the ability of the regression model to reliably learn the desired evaluator from latent representations. Second, future work can explore more diverse and application-driven characteristics. For example, quantities such as the number of holes, the minimum angle between members at joints, or other measures of geometric complexity, manufacturability should also be compatible with the TopoCtrl framework as long as they can be evaluated and labeled for training. Extending the method in this direction would further increase its usefulness as a practical design interface for engineers, especially in cases where the target quantity is easy to measure but difficult to optimize for directly. Finally, an important long-term direction is extension to three-dimensional topology optimization. Once a 3D topology foundation model becomes available, the same regression-guided latent control pipeline can be generalized to volumetric designs, enabling post-optimization manipulation of 3D structural characteristics under complex boundary conditions. This would open the door to characteristic control for practical 3D engineering components and broaden the applicability of TopoCtrl beyond planar topologies.

\FloatBarrier

\section*{ACKNOWLEDGEMENT}
The authors would like to thank the support of Tata Steel and MIT Generative AI Impact Consortium. 

\bibliographystyle{asmems4}
\bibliography{References}
\end{document}